\documentclass[fleqn,10pt]{wlscirep}
\usepackage[utf8]{inputenc}
\usepackage[T1]{fontenc}
 \usepackage{braket}
\title{Low noise all-fiber amplification of a coherent supercontinuum at 2 $\mu$m and its limits imposed by polarization noise}

\author[1,*]{Alexander M. Heidt}
\author[2]{Joanna Modupeh Hodasi}
\author[3,4]{Anupamaa Rampur}
\author [1]{Dirk-Mathys Spangenberg}
\author [1]{Manuel Ryser}
\author[3]{Mariusz Klimczak}
\author[1]{Thomas Feurer}
\affil[1]{Institute of Applied Physics, University of Bern, Sidlerstrasse 5, 3012 Bern, Switzerland}
\affil[2]{Department of Physics, University of Ghana, Legon, Accra, Ghana}
\affil[3]{Glass Department, Institute of Electronic Materials Technology, Wolczynska 133, 01-919 Warsaw, Poland}
\affil[4]{Faculty of Physics, University of Warsaw, Pasteura 5, 02-093 Warsaw, Poland}

\affil[*]{Corresponding author: alexander.heidt@iap.unibe.ch}

\begin{abstract}
We report a low noise, broadband, ultrafast Thulium / Holmium co-doped all-fiber chirped pulse amplifier, seeded by an Erbium-fiber system spectrally broadened via coherent supercontinuum generation in an all-normal dispersion photonic crystal fiber. The amplifier supports a -20~dB bandwidth of more than 300~nm and delivers high quality 66~fs pulses with more than 70~kW peak power directly from the output fiber. The total relative intensity noise (RIN) integrated from 10 Hz - 20 MHz is 0.07\%, which to our knowledge is the lowest reported RIN for wideband ultrafast amplifiers operating at 2 $\mu$m to date. This is achieved by eliminating noise-sensitive anomalous dispersion nonlinear dynamics from the spectral broadening stage. In addition, we identify the origin of the remaining excess RIN as polarization modulational instability (PMI), and propose a route towards complete elimination of this excess noise. Hence, our work paves the way for a next generation of ultra-low noise frequency combs and ultrashort pulse sources in the 2 $\mu$m spectral region that rival or even outperform the excellent noise characteristics of Erbium-fiber technology.  
\end{abstract}
\begin{document}

\flushbottom
\maketitle
%
%
\thispagestyle{empty}

\section*{Introduction}

Ultrashort pulse sources and frequency combs operating in the 2 $\mu$m spectral region have become indispensable tools for spectroscopy and precision metrology, as well as important stepping stones for the exploration of the molecular fingerprint region in the mid-infrared via nonlinear frequency conversion \cite{Schliesser2012, Muraviev2018, Lee2017, Gaida2018, Butler2019}. The application-driven requirements for such sources are constantly becoming more demanding, and include wide spectral bandwidth, ultra-low noise, and high robustness. Direct laser emission in the 2 $\mu$m wavelength regime is possible through Thulium (Tm)- or Holmium (Ho)-doped silica fiber lasers, which have been scaled to the kilowatt average power level even in ultrafast operation, but with limited spectral bandwidths \cite{Gaida2018a}. Spectrally broadband sources, which take advantage of the very broad gain bandwidth provided by Tm- and Ho-ions \cite{Hemming2014, Li2016}, are often based on Tm- and Ho-doped fiber amplifiers seeded by modelocked Erbium (Er)-fiber systems, which are spectrally broadened to the 2 $\mu$m region via the generation of Raman-shifted solitons, four-wave mixing, or supercontinuum generation in the anomalous dispersion region of a highly nonlinear fiber (HNLF) \cite{Adler2012, Kumkar2012, Coluccelli2013, Hoogland2013, Coluccelli2014, Tan2016, Sobon2018, Gaida2018}. In this way, the maturity, robustness, and excellent performance parameters of Er:fiber technology operating in the 1.55 $\mu$m window can be conveniently extended to longer wavelengths and are easily power-scaled. 

However, this approach is susceptible to possible degradation of temporal coherence and excessive noise amplification due to the nonlinear processes involved in the spectral broadening in the HNLF. It is well known that the nonlinear dynamics in the anomalous dispersion regime suffer from high sensitivity to quantum noise, leading to white-noise fluctuations of amplitude and phase of the spectrally broadened seed signal usually associated with modulational instability (MI) \cite{Dudley2006, Newbury2003, Corwin2003, Newbury2007}. Consequently, the noise performance of the 2 $\mu$m system is inherently degraded in comparison to the Er-fiber seed due the introduction of excess amplitude and phase noise at the nonlinear broadening stage, which is then further increased in the subsequent amplifier. The reported integrated root mean squared (rms) relative intensity noise (RIN) values of such systems operating at 2$\mu$m are typically in the range 0.3 - 0.7\%, which is about an order of magnitude higher than corresponding Er-fiber systems operating at 1.55 $\mu$m \cite{Coluccelli2014, Coluccelli2013, Adler2012, Gaida2018, Hoogland2013}. In fact, such white-noise-like RIN originating from the nonlinear processes in the spectral broadening stage has recently been identified as the major performance limiting factor in the further development of high power frequency comb sources at 2 $\mu$m \cite{Adler2012, Gaida2018}. 

Other system designs, which directly amplify modelocked oscillators at 2 $\mu$m, exploit nonlinear compression schemes such as soliton self-compression, which are based on anomalous dispersion nonlinear effects in the amplifier fibers or subsequent nonlinear fibers to reach comparable spectral bandwidths and pulse durations as the Er-fiber seeded systems. Hence, they are similarly susceptible to nonlinear noise amplification with reported integrated RIN values of > 0.3\%\cite{Heuermann2018, Nomura2017, Gaida2015}.

In this work we overcome these current limitations by completely eliminating noise-sensitive anomalous dispersion nonlinear dynamics from the spectral broadening stage. Instead, we employ an all-normal dispersion photonic crystal fiber (ANDi PCF) for coherent supercontinuum (SC) generation \cite{Heidt2010, Heidt2016}, which serves as the broadband coherent seed of a Tm/Ho amplifier. Normal dispersion nonlinear dynamics were recently reported to have a factor of 50 higher threshold for the onset of noise-driven decoherence compared to equivalent spectral broadening in the anomalous dispersion regime \cite{Heidt2017}. In addition, all-normal dispersion nonlinear processes also minimize the temporal jitter between different wavelength components \cite{Rothhardt2012a}, can exhibit lower RIN than the pump laser \cite{rao2019, Genier2019}, and produce flat, smooth and stable spectra that can be compressed to high quality single-cycle pulses \cite{Heidt2011b, Demmler2011}. Hence, ANDi SC are ideally suited for the coherent seeding of ultra-low noise, broadband, ultrafast fiber amplifier systems. While carrier-envelope stabilized ANDi SC pulses were previously used for seeding of ultra-broadband optical parametric amplifiers \cite{Rothhardt2012b}, which were instrumental in enabling the first demonstrations of high harmonics generation and isolated attosecond pulses at high average power \cite{Rothhardt2017}, this approach has so far not been applied to fiber optic systems. 

We report a broadband, ultrafast Tm/Ho co-doped all-fiber chirped pulse amplifier, seeded by an Er-fiber system spectrally broadened in an ANDi PCF, supporting a -20~dB bandwidth of more than 300~nm and delivering high quality 66~fs pulses with more than 70~kW peak power directly from the output fiber. This is amongst the shortest reported pulse durations for similar systems that do not use nonlinear post-compression schemes \cite{Coluccelli2013, Hoogland2013, Sobon2018, Adler2012, Kumkar2012, Tan2016}. The total RIN integrated from 10 Hz - 20 MHz is 0.07\%, which to our knowledge is the lowest reported RIN for wideband ultrafast amplifiers operating at 2 $\mu$m to date. The value further reduces to 0.03\% if the more commonly quoted range up to 1 MHz is considered. In addition, we identify the origin of the remaining excess RIN as polarization modulational instability (PMI), and propose a route towards complete elimination of this excess noise. Hence, our work paves the way for a next generation of ultra-low noise frequency combs and ultrashort pulse sources in the 2 $\mu$m spectral region that rival or even outperform the excellent noise characteristics of Er-fiber technology.

\section*{System Design and Results}

\begin{figure}[htb]
\centering
\includegraphics[width=0.8\linewidth]{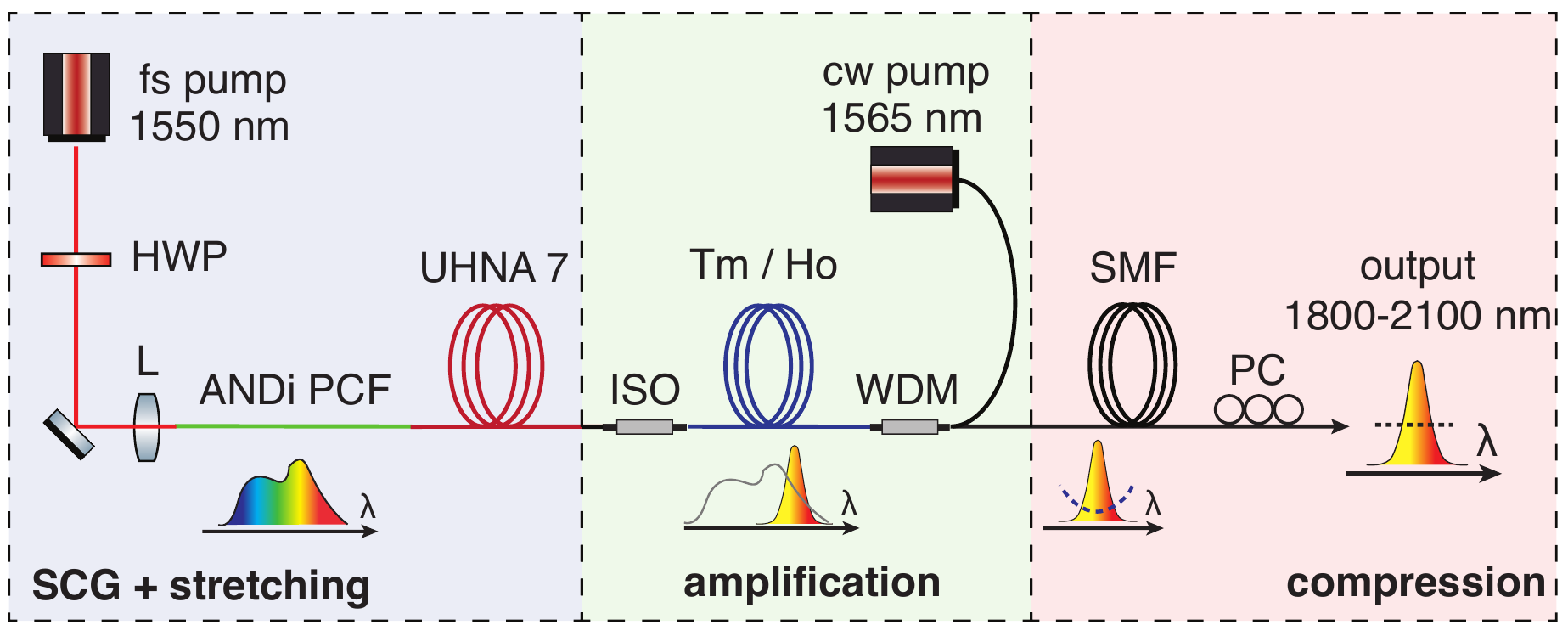}
\caption{Layout of the fiber chirped pulse amplifier. The evolution of the spectrum is schematically illustrated below the fiber sections. Dashed lines in the spectrum symbolize spectral phase. HWP - half-wave plate; L - aspheric lens; ISO - isolator; WDM - wavelength division multiplexer; PC - polarization controller.}
\label{fig:setup}
\end{figure}

The experimental setup is illustrated in Fig. \ref{fig:setup}. We employ an fiber chirped pulse amplifier (FCPA) scheme, consisting of a spectral broadening stage based on ANDi SC generation, followed by a dispersive pulse stretching and Tm/Ho co-doped fiber amplifier stage, and finally a fiberized pulse compression stage. While the Er-fiber seed laser pulses are free-space coupled to the ANDi PCF, the subsequent FCPA system is realised in an all-fiber configuration such that the amplified and compressed pulses are directly available at the fiber exit of the system.

\begin{figure}[htb]
\centering
\includegraphics[width=0.8\linewidth]{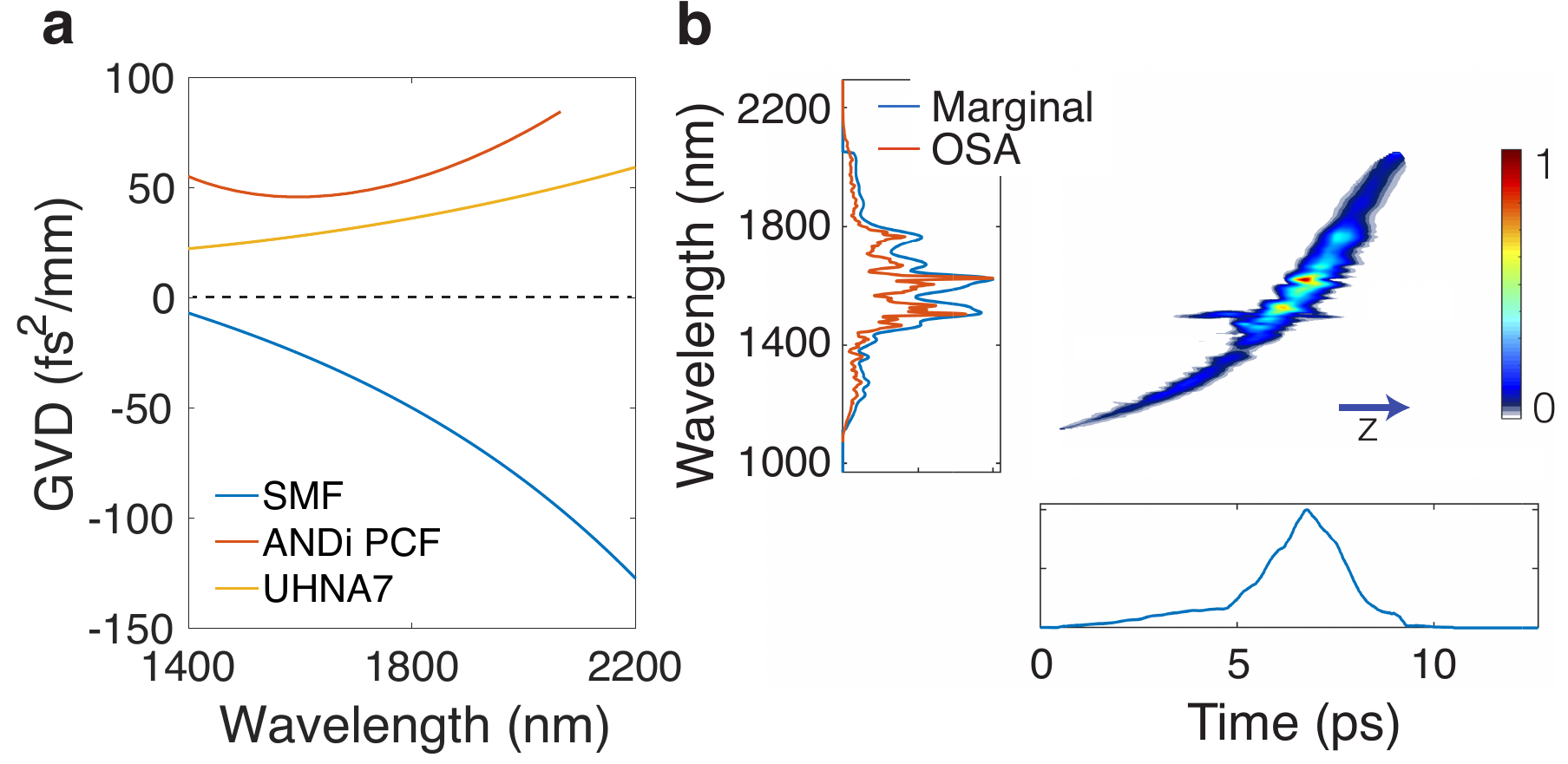}
\caption{(a) Measured group velocity dispersion of ANDi PCF, UHNA~7 and standard single mode fiber (SMF) \cite{Klimczak2016, Ciacka2018}. (b) Measured projected axis spectrogram of the SC pulse in linear scale at the exit of 20~cm ANDi PCF under similar pumping conditions as used in the amplifier system. The correlation of the spectral marginal to an independently measured optical spectrum analyzer (OSA) trace validates the accuracy of the measurement. $z$ indicates the propagation direction.}
\label{fig:properties}
\end{figure}

\begin{figure}[htb]
\centering
\includegraphics[width=0.8\linewidth]{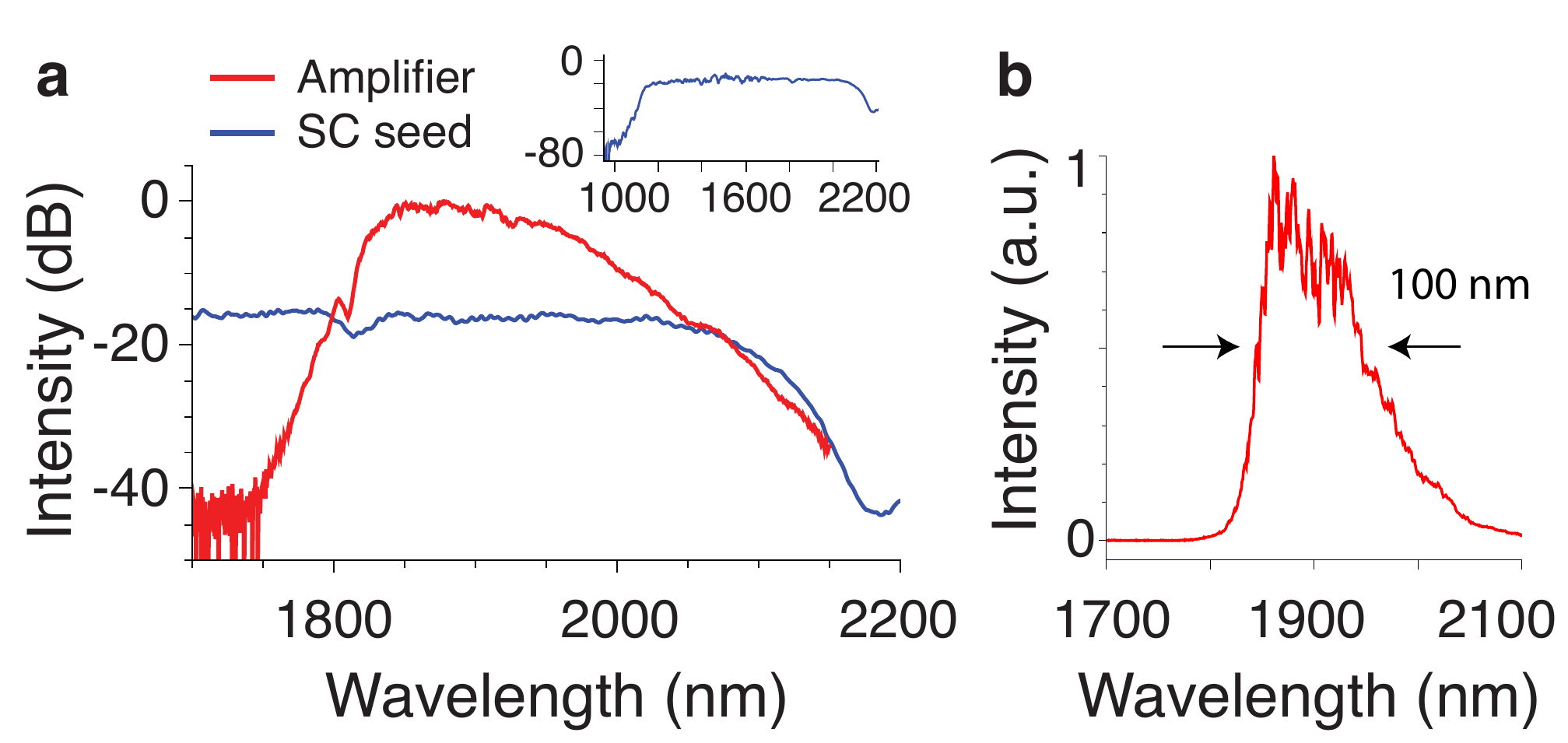}
\caption{(a) Logarithmic spectra of SC seed pulse and amplifier (not to scale). The inset shows the complete spectrum of the spectrally broadened seed pulse. (b) Linear amplifier spectrum.}
\label{fig:spectra}
\end{figure}

The seed pulses are provided by a commercial Er-fiber system, delivering 100~fs pulses with 80~MHz repetition rate at a central wavelength of 1560~nm. Coherent SC generation in a 20~cm long in-house drawn ANDi PCF is used to spectrally broaden the pulses sufficiently for seeding the entire gain bandwidth of the Tm/Ho amplifier in the 1750 - 2200 nm range. The coupled peak power is estimated to 20~kW. The fiber provides high nonlinearity ($\gamma \approx$ 214 (W km)$^{-1}$ at 1560~nm) and normal group velocity dispersion (GVD) over the entire wavelength region of interest, as shown in the measured GVD curve in Fig. \ref{fig:properties} (a). The full SC spectrum at the exit of the ANDi PCF, shown in logarithmic scale in the inset of Fig. \ref{fig:spectra} (a), covers the wavelength range 1120 - 2150~nm at a -20~dB level. Figure \ref{fig:properties} (b) shows a typical measured projected axes spectrogram of the SC pulse in linear scale at the exit of the ANDi PCF under similar pumping conditions as used in the FCPA. It is evident that the normal dispersion nonlinear processes, dominated by self-phase modulation and optical wave braking, preserve the temporal integrity of the SC pulse \cite{Heidt2017}. The chirp is predominantly linear at the long wavelength wing between 1750 - 2200~nm, i.e. the SC pulse is very well suited as a seed for the ultrafast Tm/Ho amplifier.

The SC pulses are further dispersively stretched in 6.8~m of UHNA~7 fiber (Nufern; core diameter 2.4~$\mu$m, 0.41~NA), which has normal dispersion in the Tm/Ho amplification window and is used to compensate the anomalous dispersion of the active and passive fibers of the subsequent amplifier system \cite{Ciacka2018}. The measured GVD curve of this fiber is shown in Fig. \ref{fig:properties} (a) and compared to the GVD of the standard single-mode fiber (SMF) (SM 2000, Thorlabs) used as pigtail fiber of the passive components in the amplifier. Note that we did not measure the GVD of the active fiber used in our system, but based on previous reports we assume it to be similar to passive SMF \cite{Tang2015}. 
The measured dispersion coefficients of UHNA~7 at 1900~nm ($\beta_2 = +41~$fs$^2$/mm, $\beta_3 = -99~$fs$^3$/mm) compensate not only second, but also the third order dispersion of SMF ( $\beta_2 = -65~$fs$^2$/mm, $\beta_3 = 241~$fs$^3$/mm).  This is an important requirement for achieving a high quality compressed pulse at the output of the system. In addition, UHNA~7 can also act as an efficient mode field adapter between ANDi PCF and SMF, because its small core with high NA efficiently collects the light exiting the PCF while the expansion of the core under fusion splicing can be used to minimize splice loss to SMF to < 0.2 dB. This allows seeding the subsequent amplifier with approx. 11~mW of average power contained in the relevant spectral range of 1750-2200~nm of the SC pulse.

The amplifier consists of 140 cm of Tm/Ho co-doped single-mode fiber (CorActive Th512; 9~$\mu$m core diameter, 0.16 NA, 150~dB/m core absorption at 790~nm), backward core-pumped via a wavelength-division multiplexer by an in-house built Erbium/Ytterbium (Er/Yb) co-doped single-mode fiber laser operating at 1560~nm. The length of the Tm/Ho doped fiber was optimized for maximum FWHM spectral bandwidth of the amplified signal, which does not coincide with the length required for operation with highest optical-to-optical efficiency. The pulses temporally compress during propagation in the amplifier due to the anomalous dispersion of the active and passive fibers, but it remains positively chirped with a pulse duration of about 6~ps at the exit of the amplifier. The final temporal recompression is then simply achieved by splicing an appropriate length (about 1.2~m) of SMF to the exit of the amplifier. Spurious reflections inside the amplifier are avoided by an in-line isolator at the amplifier input and an angle-cleaved fiber end-facet at the exit of the system. The polarization state is controlled by a half-wave plate at the input and a polarization controller near the output of the system. The seeding level is sufficient to operate the amplifier system near saturation, delivering a maximum of 0.5~W average output power at a pump power of 2~W. Further power scaling is limited by the small mode-field diameter of the current SMF compression fiber, which leads to excessive nonlinear effects and related pulse distortions at higher peak powers. However, this can be easily circumvented by using large-mode area compression fibers or external bulk compressors.

The spectrum of the amplified signal in relation to the ANDi SC seed is shown in Fig.~\ref{fig:spectra}. The SC spectrum is sufficiently broad to coherently seed the entire gain bandwidth of the amplifier. As a consequence, incoherent amplified spontaneous emission (ASE) noise is not observed in the spectral domain, and is also not expected to build up temporally between pulses due to the high repetition rate (80~MHz) in comparison to the long ($\sim$ ms) excited state life time of Tm- and Ho-ions. The amplified signal has a central wavelength of 1900~nm, a -20~dB spectral bandwidth of 310~nm (1798-2108~nm) and a FWHM of 100~nm. The Fourier-limited pulse duration is 45 fs, obtained from direct Fourier-transformation of the measured spectrum assuming flat spectral phase, resulting in a time-bandwidth product of TBP = 0.375. This indicates a pulse shape half way in between Gaussian and $sech^2$. Note that we do not apply any nonlinear pulse compression scheme in or after the amplifier, such as soliton self-compression, as was done in some previous demonstrations of similar amplifier systems \cite{Gaida2015, Gaida2018}.

\begin{figure}[htb]
\centering
\includegraphics[width=0.8\linewidth]{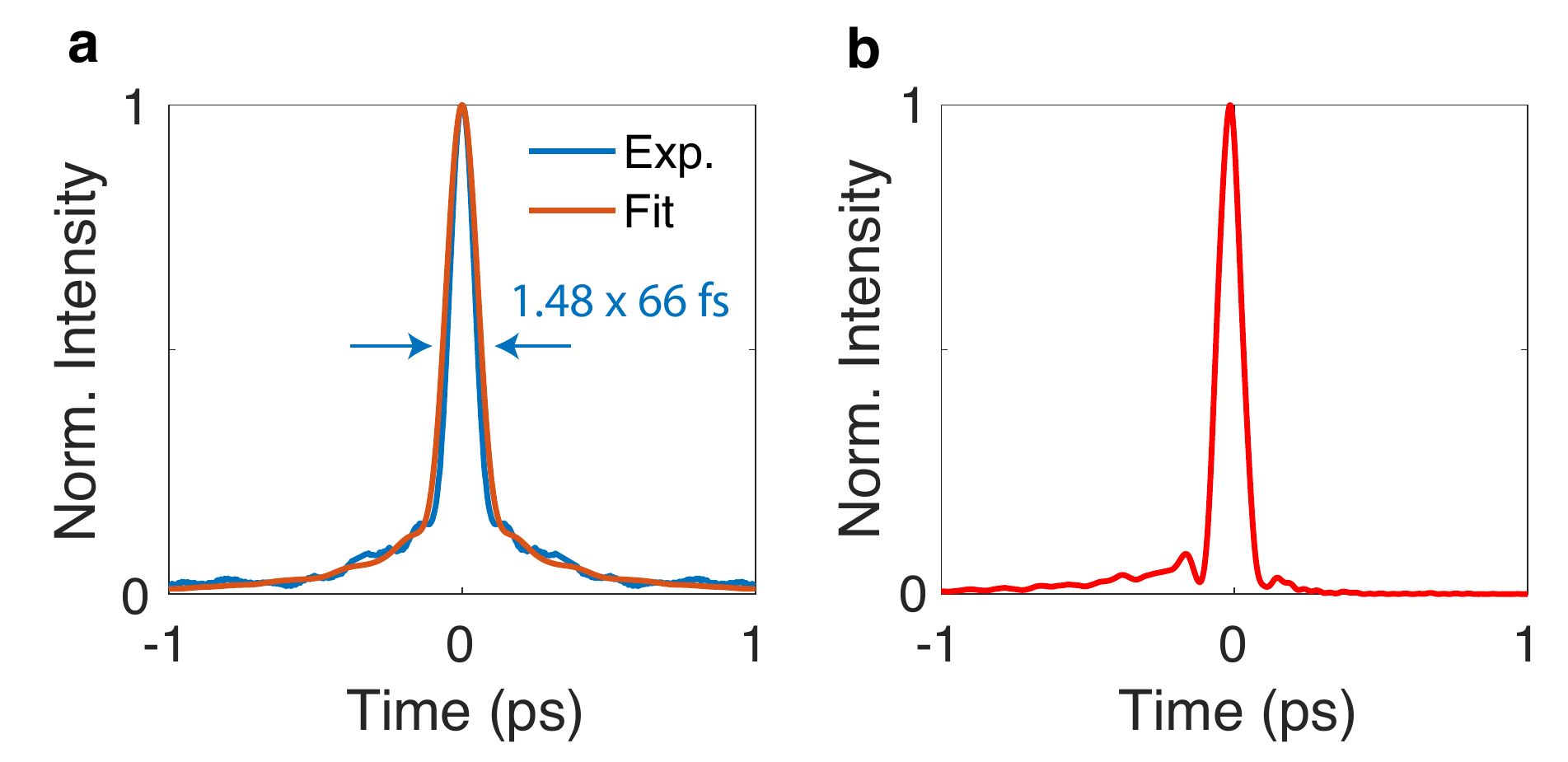}
\caption{(a) Measured autocorrelation trace in comparison to the calculated trace obtained by a procedure described in Methods. (b) Deconvoluted pulse shape of the calculated trace.}
\label{fig:ac}
\end{figure}

Figure~\ref{fig:ac}~(a) shows the measured autocorrelation (AC) trace of the system output with a FWHM of 98~fs. Applying the same deconvolution factor as obtained for the Fourier-limited pulse results in a pulse duration of 66~fs. The AC trace is exceptionally clean for an all-fiber CPA system; it does not exhibit any major side peaks, only the broadened base of the main peak indicates the presence of a small pedestal. The absence of further side pulses was checked using an AC scan window of 150~ps as well as a fast photodiode and oscilloscope combination with 60~ps temporal resolution. The approximate deconvoluted pulse shape in Fig. \ref{fig:ac}, obtained computationally by an iterative procedure reproducing the measured AC trace (see Methods), reveals that 75\% of the energy is contained in the central peak, with the remaining part mostly located in a low-level pedestal at the leading pulse edge. Using the measured pulse energy and duration and correcting for the energy content of the pedestal, we estimate the peak power of the output pulse to approximately 71~kW.


\begin{figure}[htb]
\centering
\includegraphics[width=0.66\linewidth]{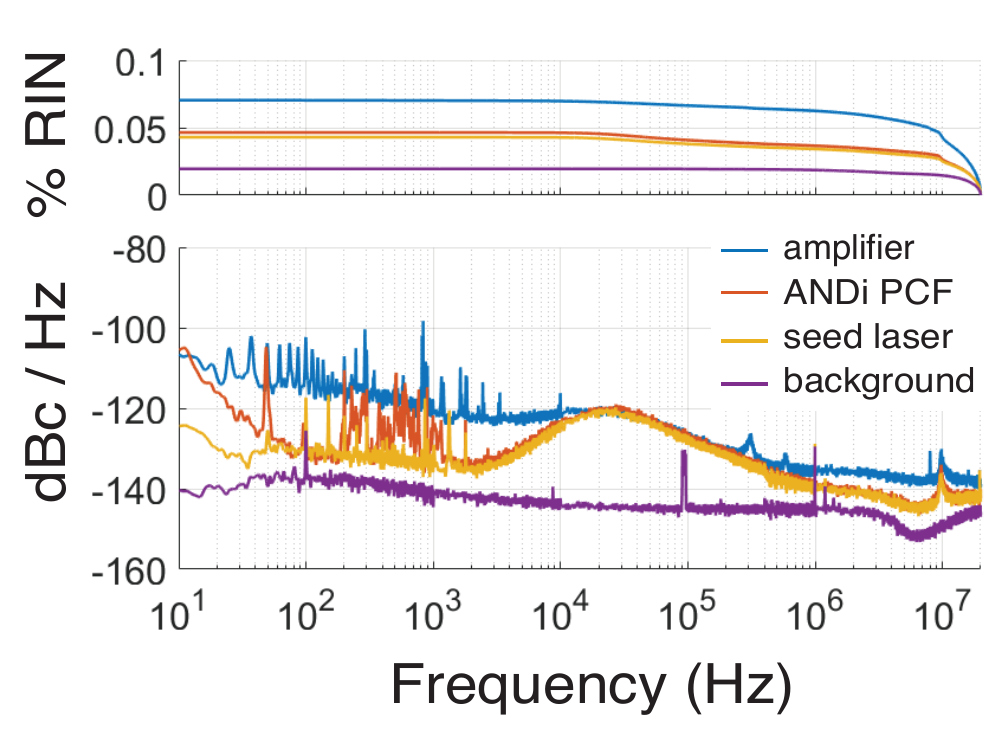}
\caption{Amplitude noise measurements at different positions along the amplifier. The bottom shows the spectrally resolved amplitude noise in a range 10 Hz - 20 MHz, while the top shows the total integrated RIN in percent. Shown are the noise of the Er-fiber seed laser before spectral broadening (yellow), the spectrally broadened seed after the ANDi PCF (red), and the amplified and compressed output of the Tm/Ho amplifier (blue). The noise floor of the measurement system is shown in purple.}
\label{fig:noise}
\end{figure}

Fig. \ref{fig:noise} shows the amplitude noise spectrum as well as the integrated rms RIN in the range 10 Hz - 20 MHz measured at different positions along the amplifier. The RIN of the SC measured directly after the ANDi PCF (0.047\%) is virtually identical to the RIN of the Er-fiber seed laser (0.045\%), except for additional noise contributions in the frequency range 200 Hz - 1 kHz. In order to facilitate other experiments, the Er-fiber system was located on a different optical table than the ANDi PCF and the rest of the experiment, requiring a free space beam path of about 3 meters length between the exit of the Er-fiber system and the coupling stage of the ANDi PCF. Hence, we attribute these additional low frequency noise components to vibrations caused by the laboratory environment resulting in beam pointing instabilities, which are translated to amplitude noise at the PCF input. After the amplifier, the total integrated RIN increases to 0.07\%. If the more commonly quoted range up to 1 MHz is considered, the RIN of the amplifier output is 0.03\%. Although this is an excellent value, we do observe an increase by a factor of about 1.5 compared to the RIN measured directly from the Er-fiber seed system, which is caused predominantly by an additional broadband noise component in the range 500 kHz - 20 MHz. In the medium frequency range 20 - 300 kHz the amplifier RIN is dominated by the RIN of the Er-fiber seed system. We have determined that the excess noise of the amplifier in the low frequency range < 20 KHz originates from the continuous wave Er/Yb pump laser of the amplifier, more specifically from the electronic driver of the 980 nm semiconductor pump diodes employed in this laser. However, this low frequency noise has only negligible influence on the integrated RIN of the amplifier.

\begin{figure}[htb]
\centering
\includegraphics[width=\linewidth]{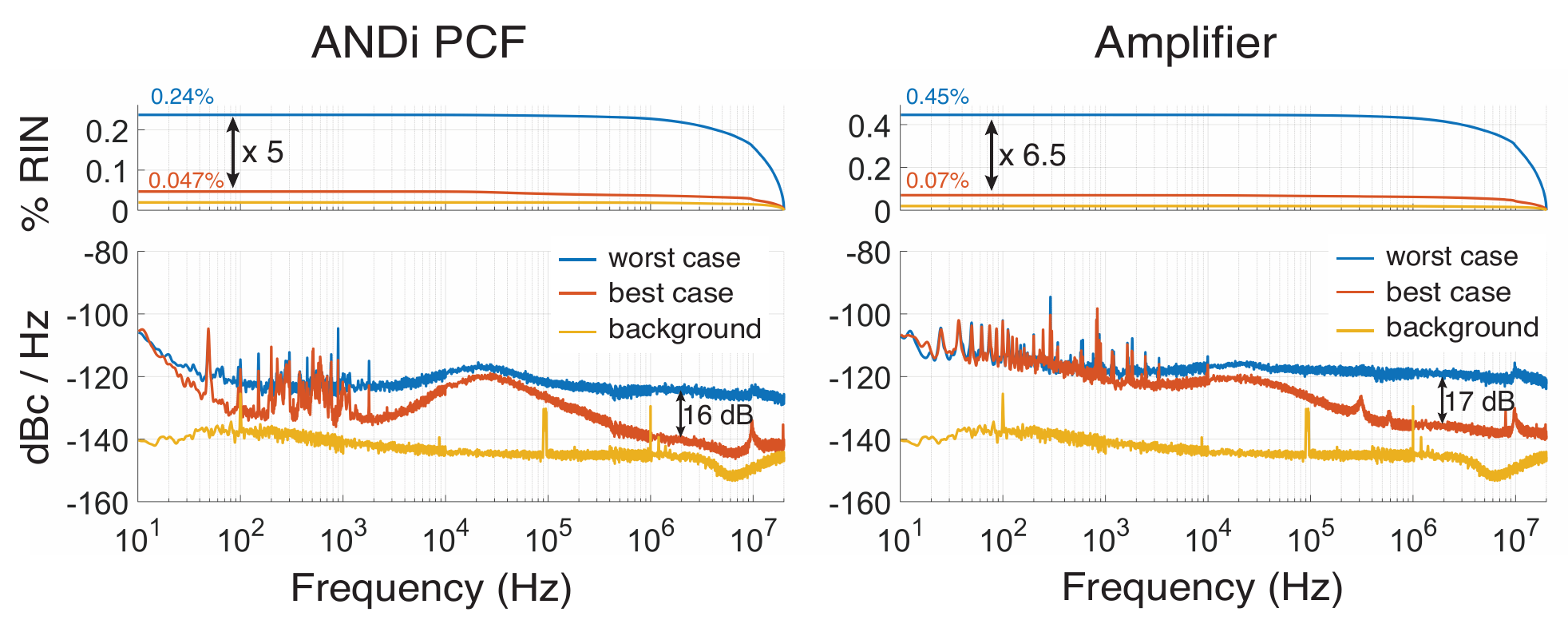}
\caption{Dependence of the measured amplitude noise on the polarization state of the Er-fiber seed pulses coupled into the system. Best and worst case RIN measurements of the spectrally broadened seed after the ANDi PCF (left) and the amplifier output (right) obtained by rotating the half-wave plate at the input of the FCPA.}
\label{fig:pol}
\end{figure}

 We observe a strong and sensitive dependence of the RIN on the input polarization state of the Er-fiber seed pulses coupled into the FCPA. This is illustrated in Fig. \ref{fig:pol}, both for the RIN measured directly after the ANDi PCF and after the amplifier. The input polarization of the seed pulses entering the ANDi PCF is varied by rotating the half-wave plate placed into the free-space beam path, and the best-case RIN measurements that were presented in Fig. \ref{fig:noise} are compared to the observed worst-case RIN measurements obtained for a different setting of the half-wave plate. For the spectrally broadened seed after the ANDi PCF, in the worst-case setting the integrated RIN increases to 0.24\% or a 5-fold increase compared to the best-case setting. This increase is caused by an additional white-noise component that raises the entire amplitude noise spectrum by about 16~dB in the high frequency range. A similar behaviour is observed after the amplifier with a 6.5-fold increase of integrated RIN to 0.45\% in the worst-case scenario. While the total RIN oscillates between these extrema in a complex pattern when rotating the input polarization, whose detailed analysis will be subject to future work, we observe only one stable global maximum and minimum setting. These extrema are very sensitive to the input polarization angle, with the RIN changing rapidly in a $\pm 1$ degree range around them. The optimum input polarization angle changes with environmental conditions such as temperature or fiber coiling radius.

\section*{Discussion}

To our knowledge, the low amplifier RIN of 0.07\% (10 Hz - 20 MHz) and the small noise amplification factor of 1.5 represent the lowest values reported for wideband ultrafast amplifiers operating at 2 $\mu$m to date. Our amplifier system is designed in such a way that many of the traditional linear and nonlinear noise amplifying effects are effectively suppressed. Firstly, the normal dispersion nonlinear spectral broadening dynamics in the ANDI PCF exclude soliton dynamics and MI, which are associated with a loss of temporal coherence and corresponding white-noise amplitude and phase instabilities occurring in fibers with anomalous dispersion. Secondly, the chirped pulse amplification scheme and avoidance of nonlinear self-compression in the amplifier minimize nonlinear noise-amplifying effects in the amplifier fibers, which do have anomalous dispersion at the operating wavelength. Thirdly, the broad spectral bandwidth of the ANDi SC coherently seeds the entire gain bandwidth of the Tm/Ho-doped amplifier fiber, which in combination with the high repetition rate effectively suppresses incoherent ASE noise. These measures explain the excellent performance of our amplifier in comparison to previous reports, where much larger RIN values and noise amplification factors were observed \cite{Adler2012, Hoogland2013, Coluccelli2013, Coluccelli2014, Gaida2018}. 

Nevertheless, we do observe highly polarization sensitive excess RIN both after the ANDi PCF and the amplifier largely caused by white-noise-like instabilities. At the pump pulse durations we employ during SC generation ($\sim$ 100 fs), the only noise-amplifying nonlinear effect occurring in ANDi fibers is polarization modulational instability (PMI) \cite{Gonzalo2018, Wabnitz1988}. In fact, the strong polarization dependence of the RIN clearly identifies PMI as source of the excess noise because the PMI gain is strongly sensitive on the relative orientation between input polarization state and the fiber's principal axes \cite{Wabnitz1988} . PMI leads to significant pulse-to-pulse fluctuations and coherence degradation of SC pulses, in particular in weakly birefringent, non-polarization-maintaining fibers as we use in this system \cite{Gonzalo2018}. As unseeded PMI can be understood as the amplification of quantum or shot noise with random phase, it appears in the noise spectrum as white-noise instability \cite{Liu2015}. This explains the experimentally observed RIN variations with rotation of the input polarization. Since we use non-polarization-maintaining fibers, the principal axes of the fibers are not well-defined and originate from an unintentional and weak birefringence induced by stress, e.g. from the drawing process or coiling on the optical table. Hence, the orientation of the principal axes may change with environmental conditions, as observed experimentally. While the ANDi PCF should be the main source of PMI noise due to the larger nonlinearity and high peak power of the propagating pulses, it is reasonable to assume additional PMI noise being generated in the amplifier as well. This is especially true if high peak powers are reached in the amplifier due to the use of nonlinear pulse shortening mechanisms, such as soliton self-compression, as was the case in recent reports \cite{Gaida2015, Gaida2018, Heuermann2018}. 

It is important to note that PMI is not a phenomenon restricted to normal dispersion fibers, but also occurs in anomalous dispersion fibers in addition to the more well-known noise amplifying effects \cite{Wabnitz1988}. We would like to point out that PMI might be an important performance limiting factor contributing to the white-noise-like amplitude and phase instabilities observed in recent demonstrations of high power frequency combs at 2 $\mu$m, which were seeded by SC pumped in the anomalous dispersion regime and realized with non-polarization-maintaining fibers \cite{Gaida2018}.

PMI in the SC generation process can be suppressed by employing only highly birefringent, polarization-maintaining (PM) fibers pumped along their principal axis \cite{Wabnitz1988, Gonzalo2018, Liu2015}. In this case, the dominating noise-amplifying nonlinear effect in normal dispersion SC dynamics is the nonlinear coupling of stimulated Raman scattering and parametric four-wave mixing \cite{Heidt2017}.  At the peak powers typically used for SC generation, this effect occurs only for pump pulse durations exceeding 1~ps and is effectively suppressed for shorter pump pulses \cite{Heidt2017}. Hence, our work paves the way for the construction of next generation ultra-low noise high power ultrashort pulse amplifiers and frequency combs practically free of excess noise by simply replacing all the fibers in our system with their corresponding PM version. Since PM ANDi fibers and corresponding SC sources are now available \cite{Dobrakowski2019, Tarnowski2019}, this is a straightforward task. Indeed, our preliminary tests with a PM ANDi SC source and an all-PM chirped pulse amplification system indicate that the RIN after the amplifier is very close to the RIN of the Er-fiber seed system \cite{Rampur2019}. 

\section*{Methods}
\subsection*{ANDi PCF}
The fiber was drawn in-house with a core diameter of 2.3~$\mu$m surrounded by a photonic crystal lattice realized in an all-solid design using two thermally compatible silicate glasses. Schott SF6 is used as core and lattice glass, Schott F2 for lattice inclusions and tube glass. More details to fiber design and production are given elsewhere \cite{Klimczak2017}.

\subsection*{Deconvolution of the measured autocorrelation trace}
For evaluation of the pulse quality we computationally reproduce the measured AC trace by adding third and fourth order spectral phase terms to the measured spectrum, taking the Fourier transform and calculating the AC trace of the corresponding pulse. The uncompensated third order phase $\Phi_3$ is assumed to stem predominantly from linear propagation through the FCPA system. Therefore, it is calculated from the measured GVD curves in Fig. 2 (a) and the lengths of the different fiber types used in the system to approximately $\Phi_3 = 3.5 \times 10^{-4}$ ps$^3$. The fourth order spectral phase $\Phi_4$ is then determined in an iterative procedure finding the best fit to the experimentally measured trace. A good fit is found for $\Phi_4 = -3.5 \times 10^{-5}$ ps$^4$, as shown in Fig. \ref{fig:ac} (a). While the calculation slightly overestimates the pulse duration, the shape of the AC trace is well reproduced.

\subsection*{Relative intensity noise measurements}
The measurements were conducted according to the standards outlined in literature \cite{Scott2001}. The amplitude noise was analyzed by an amplified photodiode (Thorlabs PDA10D2, bandwidth DC-25 MHz, 900 - 2600 nm spectral range, 5 kV/A transimpedence gain) connected to an electronic spectrum analyzer (ESA) (Signal Hound USB-SA44B, bandwitdh 1 Hz - 4.4 GHz). A DC block capacitor with cut-off frequency < 3 Hz was used at the input of the ESA. The photodiode signal was additionally filtered by a 20~MHz low-pass filter to avoid saturation of the ESA at the pulse repetition rate. The full noise power spectrum was then recorded in multiple steps with resolution bandwidths adapted to the recorded frequency range. The chosen resolutions were: 1 Hz in the range 10 Hz - 1 kHz, 10 Hz in the range 1 - 10 kHz, 100 Hz in the range 10 kHz - 1 MHz, and 10 kHz above 1 MHz). 50 averages were taken in each section. The measured spectra are normalized by the corresponding noise-equivalent bandwidths to yield the mean electrical noise power spectrum $\langle \mathcal{P}_N (f) \rangle$, which is then related to the measured mean electrical DC power $\langle \mathcal{P}_{DC} \rangle$ to yield the $RIN (f) = \langle \mathcal{P}_N (f) \rangle / \langle \mathcal{P}_{DC} \rangle$ and displayed in logarithmic units dBc/Hz. Since $\mathcal{P}_N \propto P_N^2$ (optical noise power) and $\mathcal{P}_{DC} \propto P_C^2$ (optical carrier power), the root mean square optical intensity fluctuations are given by $\sqrt{\int RIN(f) \textnormal{d}f}$ integrated over a given frequency range. This value is used in this manuscript when integrated RIN values are referenced. 

\bibliography{olpaper}

\section*{Acknowledgements}

Swiss National Science Foundation (SNSF) (PCEFP2-181222, NCCR MUST). Fundacja na rzecz Nauki Polskiej (FNP) (First TEAM/2016-1/1). A.R. acknowledges a Fellowship by the Schlumberger Foundation (Faculty for the Future).

\section*{Author contributions statement}

A.H , M.K. and T.F. conceived the study. M.K. designed and supplied the nonlinear fiber. A.H., J.M.H., M.R. and A.R. designed and built the Tm/Ho amplifier. D.S., A.R., and A.H. performed RIN measurements and corresponding data analysis. A.R. and M.K. performed fiber dispersion measurements. A.H. and J.M.H. set up autocorrelation measurements and analysis. A.R. and M.K. performed the spectrogram measurement of the SC pulse, A.H. the analysis. All Authors contributed to scientific interpretation of results, writing, and editing of the article.
\section*{Additional information}

\textbf{Competing interests} The authors declare no competing interests. 

\end{document}